\documentclass[twocolumn,showpacs,aps,pre,groupedaddress,amssymb,amsmath]{revtex4}

\usepackage{graphicx}
\usepackage{longtable}

\begin{document}

\title{Bubble kinematics in a sheared foam}

\author{Yuhong Wang}
\author{Kapilanjan Krishan}
\author{Michael Dennin}
\affiliation{Department of Physics and Astronomy, University of
California at Irvine, Irvine, California 92697-4575}

\date{\today}

\begin{abstract}
We characterize the kinematics of bubbles in a sheared
two-dimensional foam using statistical measures. We consider the
distributions of both bubble velocities and displacements. The
results are discussed in the context of the expected behavior for
a thermal system and simulations of the bubble model. There is
general agreement between the experiments and the simulation, but
notable differences in the velocity distributions point to
interesting elements of the sheared foam not captured by prevalent
models.
\end{abstract}


\maketitle

\section{Introduction}

The microscopic kinematic response of a system to external forces
can directly relate to its macroscopic properties. While the
kinematic response of continuous media such as Newtonian fluids
has been well characterized, theoretically as well as
experimentally, understanding the behavior of non-Newtonian fluids
and materials continues to be an area of active research
\cite{LL87,BY92,C81,B00}. For both Newtonian and non-Newtonian
fluids, the particle kinematics generally is characterized by
velocity and position probability distributions
\cite{ML00,MD97,LCDKG99, OTLL03}. This is due to the large number
of particles involved and the need for an inherently statistical
description of the microscopic behavior. In studies of Newtonian
fluids, the distributions are primarily governed by thermal
fluctuations. In contrast, the fluctuations in many complex fluids
are strongly coupled to the flow and structure of the constituent
particles that are macroscopic objects, such as granular matter
and bubbles, for which the relevant energy scales are
significantly larger than thermal energies. This difference
between thermal and athermal fluctuations can lead to
qualitatively different explorations of phase space. It is
therefore imperative to characterize not only the average flow
behavior for complex fluids, but also the fluctuations of these
particles.

The fluctuations in complex fluids are governed by the microscopic
dynamics of the constituent elements of the fluid. The dynamics
are influenced by the structure and interaction at these scales.
Examples of such materials include cellular structures such as
foams \cite{WH99}, fluids containing worm-like-micelles
\cite{cddrs04,gs06} and large molecular mass solutions
\cite{MBY01}. The constraints imposed by the structure influence
the kinematic response (individual particle motions) at the
microscopic scale \cite{D04}. In aqueous foams, the local
structure consists of densely packed bubbles which elastically
deform and rearrange their configuration in response to stresses.
Such actuation occurs frequently, transitioning the system towards
the various stable structural configurations accessible to the
system. The rearrangements are strongly nonlinear and result in
spatially and temporally localized changes in the velocity of the
participating bubbles.

The kinematic response of various complex fluids has been the
subject of both experimental and theoretical work. For example,
fluctuations of particles in granular matter have been heavily
studied, especially in the context of kinematic models
\cite{MD97,M03,LCDKG99}. These fluctuations have important
implications for the macroscopic rheology as they represent
fundamental aspects of the flow at the microscopic scale.

Aqueous foams are ideal systems to investigate fluctuations
\cite{WH99}. As with granular matter, the structure has a
sufficient influence on the dynamics that the length scale of
interest is the size of the constituent bubbles rather than
molecular dimensions. In contrast to granular matter, the primary
constituent elements in such a material (air and water) are both
Newtonian fluids. The dynamics are sufficiently overdamped that
the system is effectively massless. This combination can lead to
new behavior not observed in the granular systems and leads to
interesting questions of the universality of the behavior that is
observed in a range of complex fluids \cite{D95,OTLL03}.

Many past measurements of fluctuations in foam have focused on
global quantities, such as stress fluctuations or fluctuations in
energy. A number of simulations suggest various classes of
interesting behavior for the fluctuations. Simulations for very
dry foams, using a vertex model, propose power-law scaling for the
stress fluctuations \cite{KNN89,OK95}. Quasi-static simulations of
wet foams that measured T1 events suggest power law scaling near
the melting transition \cite{WBHA92,HWB95}. (T1 events are events
in which bubbles exchange neighbors.) In contrast, a q-potts model
for foam only found evidence for power law scaling in
distributions of energy fluctuations but not in the T1 event
distributions \cite{JSSAG99}.

Another interesting model is the bubble model \cite{D95,D97}. This
treats a foam as a collection of spheres (or circles in
two-dimensions) that experience two forces: (1) a spring force
proportional to the overlap of bubbles and (2) a viscous drag
force proportional to the velocity difference between two bubbles
in contact. This model is most relevant for wet foams in which the
bubbles are essentially spherical. Simulations of the bubble model
suggest that power law scaling did not exist for energy
fluctuations \cite{D95,D97,TSDKLL99}. Instead, there is a
well-defined average size of the stress fluctuations. Various
experiments have used indirect measures of the fluctuations, and
systems with both power-law behavior \cite{KE99} and a
well-defined average \cite{GD95,DK97} have been reported. More
recently, experiments using bubble rafts have directly measured
the stress fluctuations and found behavior that is consistent with
the predictions of the bubble model \cite{LTD02,PD03}. (A bubble
raft is a model two dimensional foam consisting of bubbles
floating on a water surface \cite{BL49}.)

More recently, work has been carried out to characterize the
statistics of the individual bubble motions. A significant study
of velocity fluctuations and particle diffusion has been carried
out for the bubble model \cite{OTLL03}. One observes two distinct
regimes of behavior as a function of the applied rate of strain.
Below a critical rate of strain, a quasistatic regime is
identified that is associated with a rate of strain independent
average stress during flow. Above the critical rate of strain, the
average stress increases with increasing rate of strain. A number
of statistical measures exhibit different behavior above and below
the critical rate of strain \cite{OTLL03}. Given this relatively
complete characterization of the bubble model, it is interesting
to explore experiments that directly test these predictions. In
this regard, the bubble raft is an ideal system as it models a wet
foam in a manner that is very similar to the bubble model. By
directly comparing the experiments and the model, one can gain
insight into the usefulness of the approximations that are central
to the bubble model. In this regard, some initial experiments with
a bubble raft have tested the predicted scaling of the width of
the velocity distribution. These experiments used a highly
polydisperse system in a Couette geometry (flow between to
concentric cylinders). The measurements of the width of the
velocity distribution for this system agreed with similar
prediction from the bubble model simulations \cite{D05}.

In this paper we study the response of a bubble raft when
subjected to parallel shear. When sheared, the velocity profile of
the bubbles forming the foam are seen to asymptotically converge
to that of a Newtonian fluid. However, the fluctuations in the
velocity profiles  are seen to play an important role in
describing the impact of the inherent coarse cellular structure of
the foam. The rest of the paper is organized as follows. Section
II describes the experimental methods. Section III presents the
results, and Sec. IV provides a discussion of the results in the
context of both the simulations results in Ref.~\cite{OTLL03} and
the earlier experiments reported in Ref.~\cite{D95}. Section IV
also discusses interesting directions for future work.

\section{Experimental methods}

\begin{figure}
\includegraphics[width=8cm]{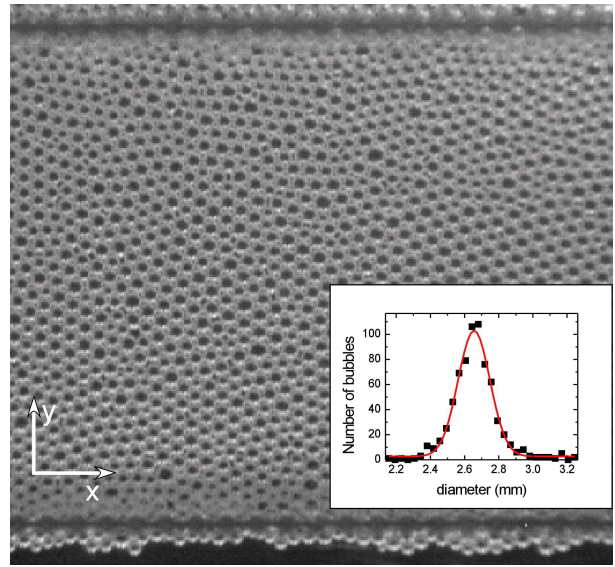}
\caption{A typical image representing the instantaneous structure
of the bubble raft. Shear induced through the bands on the top and
bottom cause the material to yield through slip between
neighboring bubbles (T1 events). These local events cause
fluctuations in the velocity profiles of the bubbles, that average
out over long times. Insert is a typical distribution for the
bubble sizes.} \label{fig_raft_image}
\end{figure}

Our experiments characterize the motion of individual bubbles of a
bubble raft when subjected to shear. The bubble raft is produced
by flowing regulated nitrogen gas through a needle into a
homogeneous solution of 80\% by volume deionized water, 15\% by
volume glycerine, and 5\% by volume Miracle Bubbles (from Imperial
Toy Corporation). The bubble size is dependent on the flow rate of
the nitrogen gas as well as the depth of the needle in the water
layer. This system forms a two-dimensional wet foam on a
homogeneous liquid substrate. The bubble raft formed consists of a
tightly packed configuration of bubbles as seen in
Fig.~\ref{fig_raft_image}. We use relatively monodisperse bubbles
in our experiments because they are reproducible and easy to
control. The typical mean bubble diameter is 2.66 mm. A typical
size distribution is shown in the insert to
Fig.~\ref{fig_raft_image}. The width of the distribution is of the
order 0.2 mm. The bubble raft is sheared by two parallel and
counter-rotating bands driven by a stepper motor. The x-direction
($\bf{\hat{i}}$) is taken parallel to the direction of motion of
the bands, and the y-direction ($\bf{\hat{j}}$) is taken
perpendicular to the direction of motion of the bands, as
indicated in Fig.~\ref{fig_raft_image}.

The motion of bubbles are recorded using a CCD camera. The frame
rate of the CCD camera is kept high enough so as to enable
identifying individual bubbles between successive images. The
digitized images are analyzed to yield the positions and
velocities of individual bubbles during shear. The experimental
setup has been previously discussed in detail in \cite{wkd05a}.

During each run, the bubbles are subjected to a total macroscopic
shear strain of $\gamma \equiv \Delta x/D = 5$, where $\Delta x$
is the total displacement the band and $D = 57\ {\rm mm}$ is the
distance between the two bands. This permits us to carry out our
experiments in a reasonable time frame constrained by the
life-time of the bubbles forming the raft when the shear rate is
low. We find this amount of strain suffices in investigating the
asymptomatic behavior of the system \cite{wkd05a}. The velocity of
the driving bands is $v_w \bf{\hat{j}}$, and the rate of strain is
given by $\dot{\gamma} = 2v_w/D$. In this setup, the limits for
the rate of strain are $10^{-3}\ {\rm s^{-1}}$ to $10^{-1}\ {\rm
s^{-1}}$.

The image analysis techniques to measure the positions of the of
the bubbles are also detailed in \cite{wkd05a}. The central
one-third of the trough are used for all measurements reported in
this paper. This region is found not to be strongly influenced by
the entry-effects at the ends of the flowing zone. The
instantaneous (for experimental purposes) velocity of a bubble is
computed by considering the distance travelled by an individual
bubble between successive frames recorded by the CCD camera. A
longer time average may also be computed by considering the
displacement of the bubbles between images recorded over the
appropriately selected longer time. When discussing spatial
dependence of the various measures, we spatially divide the bubble
raft into equally spaced bins with a width of 1.4 mm in the
y-direction. The bins are rectangular in shape as they extend over
the middle third of the system in the x-direction.

In this paper we are primarily interested in detailing the
probability distribution for the velocity and the nature of the
bubble displacements. In order to characterize these quantities,
we report on a number of different measures. First, we consider
the root mean squared deviation of the velocity from its average
value. This is done separately for the x- and y-component, and it
is given by $\delta v_i = \sqrt{<(v_i (y)-\bar{v}_i(y))^{2}>}$. In
this expression, $i$ indicates the x or y-component of the
velocity and $v_i(y)$ refers to the instantaneous velocity
component of a bubble. $\bar{v}_i(y)$ denotes the average of these
velocities in a given bin (indicated by the y-position of the bin)
over the total strain applied. The braces, $<>$, refer to an
average over all bubbles being considered. This provides a measure
of the width of the velocity distribution. We also consider the
full probability distribution for the velocity, usually for an
individual bin in the y-direction. For comparison with results
using the bubble model, we also consider the probability
distribution for the deviation of the velocity from a linear
profile $\Delta \bf{v} \equiv \bf{v} - \bf{v}_L$ where $\bf{v}_L$
is a linear profile defined by ${\bf v_L} (x) \equiv 2v_w (x/D)
{\bf \hat{j}}$. Here $-D/2 \leq x \leq D/2$, where $L$ is the
distance between the driving bands. Finally, we characterize the
bubble displacements by consider the mean square displacement of
the bubbles as a function of time ($<(\Delta x)^2>$).

\section{results}

\begin{figure}
\includegraphics[width=8cm]{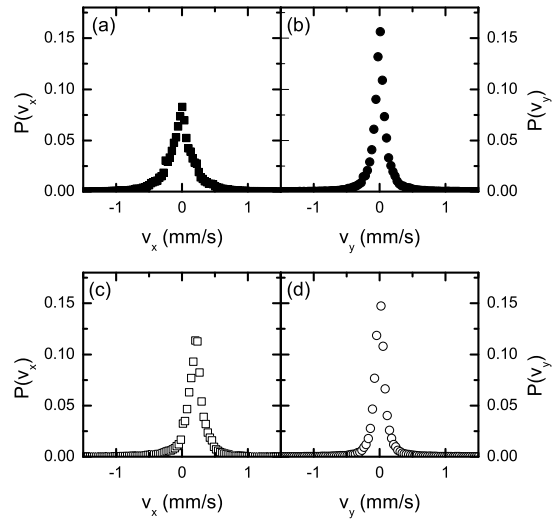}
\caption{(a) Probability distribution for $v_x$ for bubbles in the
central bin of the system. (b) Probability distribution for $v_y$
for bubbles in the central bin of the system. (c) Probability
distribution for $v_x$ for bubbles in the bin half way between the
center and the edge of the system. (b) Probability distribution
for $v_y$ for bubbles halfway between the center and the edge of
the system. All plots correspond to a shear rate of $0.0056\ {\rm
s^{-1}}$.} \label{vel-distr}
\end{figure}

The most basic characterization of the fluctuations is given by
the velocity probability distribution. We first consider the
probability distributions for different spatial bins. The results
for the center bin are presented in Fig.~\ref{vel-distr} for both
the x-component (Fig.~\ref{vel-distr}a) and y-component
(Fig.~\ref{vel-distr}b) of the velocity. As one expects, the
distributions are centered at zero for the center bins, as the
average velocity profile goes through zero in the center of the
system. Both distributions are symmetric and are well-fit by a
Lorentzian function.

Slightly different behavior is observed when considering the
distribution of velocities in an off-center bin. In this case, one
observes no fundamental change in the distribution of the
y-component (Fig.~\ref{vel-distr}d). The average $v_y$ is expected
to be zero throughout the system, and this is consistent with the
observed symmetric distribution centered at $v_y = 0$. However,
the distribution for $v_x$ is asymmetric. One measure of the
asymmetry is the third moment of the distributions. For
comparison, the distributions in Fig.~\ref{vel-distr}a-d have
third moments of $6.4 \times 10^{-5}$, $4.9 \times 10^{-4}$, $-1.4
\times 10^{-2}$, and $7.1 \times 10^{-5}$, respectively.
Therefore, the measured asymmetry for the off-center $v_x$
distribution is two to three orders of magnitude larger than any
of the other distributions. This has the interesting consequence
that the most probable value for $v_x$ is different from the
average value.

\begin{figure}
\includegraphics[width=8cm]{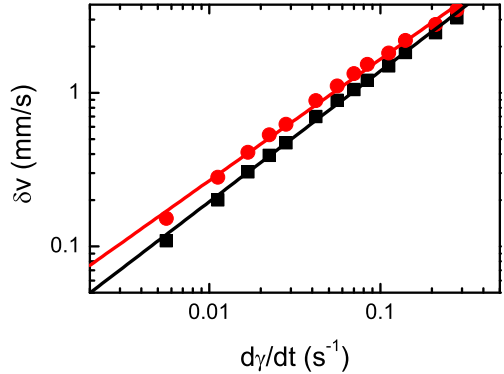}
\caption{(color online) The spread of the velocity distributions
is seen to increase with increasing strain rate as a power law.
The red circles (x-component) and black squares (y-component)
refer to the width of velocity fluctuations in a central bin of
the sheared bubbles. The solid lines are fits to the corresponding
data.} \label{width}
\end{figure}

For a thermal distribution of velocities, such as one might find
in an ideal gas, the width of the velocity distribution is related
to the temperature through the distribution for kinetic energy.
For the bubble raft, the system is highly overdamped and the
bubbles are effectively massless. Therefore, it is not clear how
one connects the velocity distribution with a temperature.
Nonetheless, various proposals exist for defining an effective
temperature as a function of the applied rate of strain,
therefore, it is useful to consider the rate of strain dependence
of the width of the velocity distribution. For direct comparison
of the results of Ref.~\cite{OTLL03} and \cite{D05}, we
characterize the width of the distributions using $\delta v_i$ for
the central bins. This choice avoids complications due to
asymmetry at bins near the edges. Both $\delta v_x$ and $\delta
v_y$ increase with increasing rate of strain. For the range of
rates of strain studied here, the dependence is consistent with a
power law with an exponent of $0.85 \pm 0.01$ for $\delta v_y$
(circles in Figure~\ref{width}) and $0.79 \pm 0.02$ (squares in
Figure~\ref{width}) for $\delta v_x$.

\begin{figure}
\includegraphics[width=8cm]{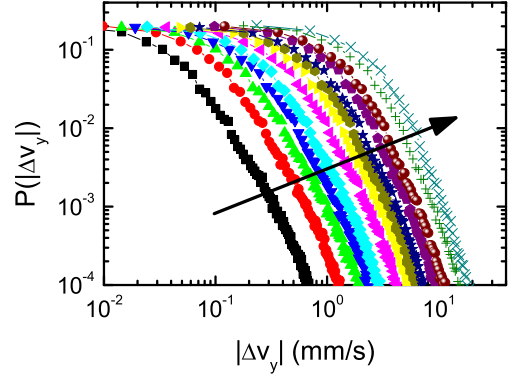}
\caption{(color online) Distribution of fluctuations in velocity
along the y-direction for all bubbles. The different symbols
(colors) represent different rates of strain, with the arrow
indicating the direction of increasing strain.} \label{velydist}
\end{figure}

Following the characterization of the fluctuations in
Ref.~\cite{OTLL03}, we also consider the probability distributions
for $\Delta v_x$ and $\Delta v_y$. This distribution is computed
over all position bins. The behavior of $\Delta v_y$ is consistent
with the expected behavior based on other characterizations (see
Fig.~\ref{velydist}). The width of the distribution increases with
increasing rate of strain, reflecting the increased flux of energy
into the system being distributed.

\begin{figure}
\includegraphics[width=8cm]{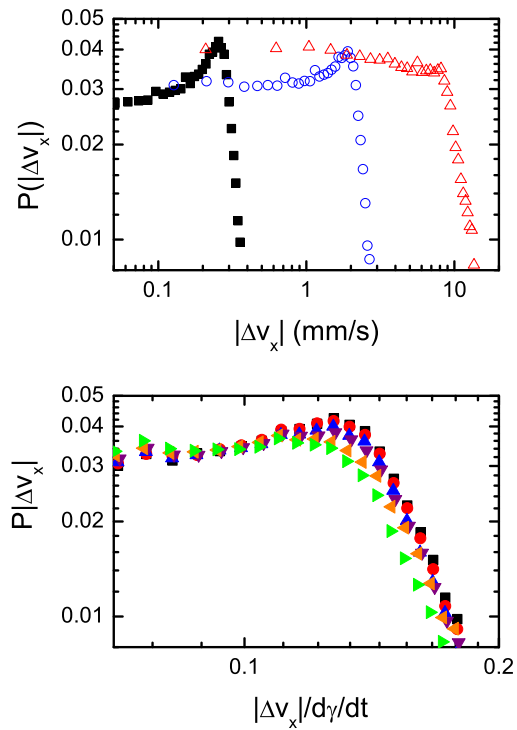}
\caption{(color online) (a) Probability distribution of
fluctuations in velocity along the x-direction ($\Delta v_x$) for
all bubbles for rates of strain of $0.0056\ {\rm s^{-1}}$
($\blacksquare$), $0.042\ {\rm s^{-1}}$ (blue $\circ$), $0.21\
{\rm s^{-1}}$ (red $\bigtriangleup$). The three curves illustrate
the existence of peak that disappears at higher rates of strain.
(b) Probability distribution of fluctuations in velocity along the
x-direction ($\Delta v_x$) for all bubbles for a range of rates of
strain. Here $\Delta v_x$ is scaled by the rate of strain. This
highlights the linear dependence of the peak on the rate of
strain.} \label{velxdist}
\end{figure}

\begin{figure}
\includegraphics[width=8cm]{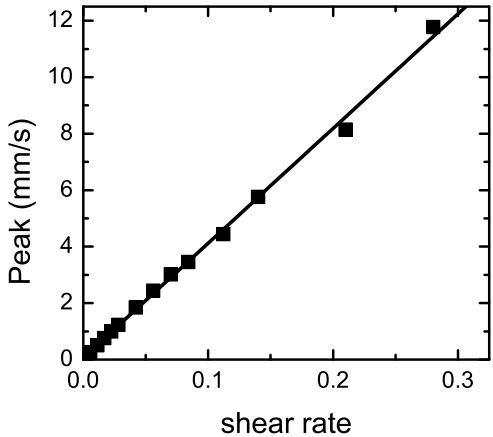}
\caption{Location of the peak in the distribution of velocity
fluctuations in the x-direction as a function of rate of strain
(symbols). The line is a linear fit.} \label{peak values}
\end{figure}

The behavior for $\Delta v_x$ indicates interesting departures
from previous studies. Again, the width of the distribution
increases with increasing rate of strain. However, there is a
well-defined peak in the distribution for slow rates of strain. As
indicated in Fig.~\ref{velxdist}a, the peak decreases in amplitude
as the rate of strain increase, and eventually disappears for
rates of strain greater than $0.1\ {\rm s^{-1}}$. The peak is
highly asymmetric, demarcating a sharp decline in probability for
larger velocity fluctuations from a gradual drop at lower ones.
The value of the peak as a function of rate of strain is plotted
in Fig.~\ref{peak values}. The scaling of this should be compared
with the width of the velocity distributions in Fig.~\ref{width}.
In contrast to the behavior of the distribution widths, the
location of the peak is consistent with a linear dependence on
rate of strain. This scaling is illustrated in
Fig.~\ref{velxdist}b where we scale the x-axis by the rate of
strain. We observe a number of features of the distribution by
doing this. First, as expected, the location of the peaks
coincide, but the amplitude of the peak is clearly decreasing.
Second, some variation is noted in the tail of the distributions,
indicating that the distributions do not scale perfectly with rate
of strain. This may be attributed to the power-law scaling of the
width of the velocity distributions.

We tested the robustness of the peak by determining this
distribution for a number of runs at the same rate of strain but
with different configurations of bubbles. Though there was some
variation in the amplitude of the peak from run to run, the peak
was a distinct feature in each run.

\begin{figure}
\includegraphics[width=8cm]{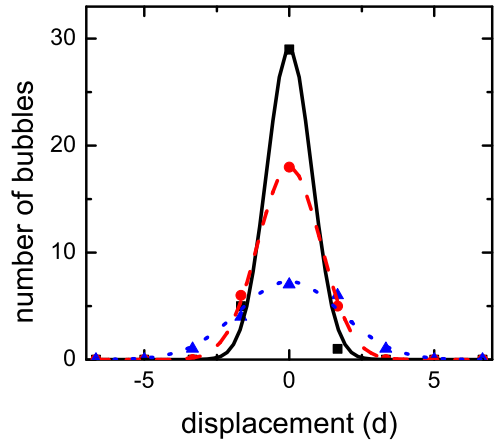}
\caption{(color online) Plotted here are the number of bubbles
that start in the central bin that experience a particular
displacement (in units of the bubble diameter $d$) in a given
time. The symbols are the data for three different times: 25 s
($\blacksquare$), 100 s (red $\bullet$), and 320 s (blue
$\blacktriangle$). The curves are fits of each set of data to a
Gaussian distribution.} \label{displacements}
\end{figure}

Velocity distributions are not directly informative on the
dynamics of individual bubbles. The second set of measures focuses
on the particle displacements. For the data presented here, we
focus exclusively on the motion in the y-direction and on bubbles
that start in the central bin. This is done to discriminate
transport induced due to the underlying constant shear along the
x-direction. The motion of these bubbles is approximately
diffusive. One measure of this is to compute the histogram of
displacements for a given starting position for a bubble. This is
shown as a function of time for a rate of strain of $0.0056\ {\rm
s^{-1}}$ in Fig.~\ref{displacements}. We observe the width of the
distribution to be increasing in time, consistent with a diffusive
process.

\begin{figure}
\includegraphics[width=8cm]{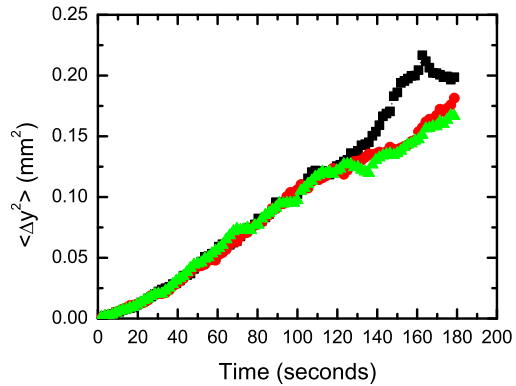}
\caption{(color online) Plotted here is $<(\Delta x(t))^2>$
averaged over all the bubbles as a function of time starting in
three different initial bins: the central bin ($\blacksquare$),
$8.55\ {\rm mm}$ from one band (red $\bullet$), and $8.55\ {\rm
mm}$ from the other band (green $\blacktriangle$). For the off
center locations, we observe the expected symmetric behavior, with
some deviation from linear at late times due to the influence of
the boundaries.} \label{diffusion}
\end{figure}

Another way to characterize this motion is to consider directly
the mean square displacement of the bubbles as a function of time
($<(\Delta x(t))^2>$). This is shown in Fig.~\ref{diffusion} for
bubbles starting from three different locations in the trough: the
central bin and two bins chosen symmetrically on either side of
the central bin. As expected for diffusion, the behavior of
$<(\Delta x(t))^2>$ is essentially linear in time in the central
bin. For the off center bins, the displacements at longer times
are are suppressed. This is presumably due to the confining
effects from the boundaries of the flow region.

An interesting feature of the data is the fact that for extremely
short times (less than 15 seconds) the behavior clearly deviates
from linear. In a system of particles in a gas, one would expect
this for time scales short enough that ballistic motion is
observed. For the bubbles, these short times presumably correspond
to the linear motion of bubbles in between the T1 events and
corresponding bubble rearrangements. We are currently pursuing
more detailed tracking of individual bubbles to determine the
exact nature of the short time behavior.

\begin{figure}
\includegraphics[width=8cm]{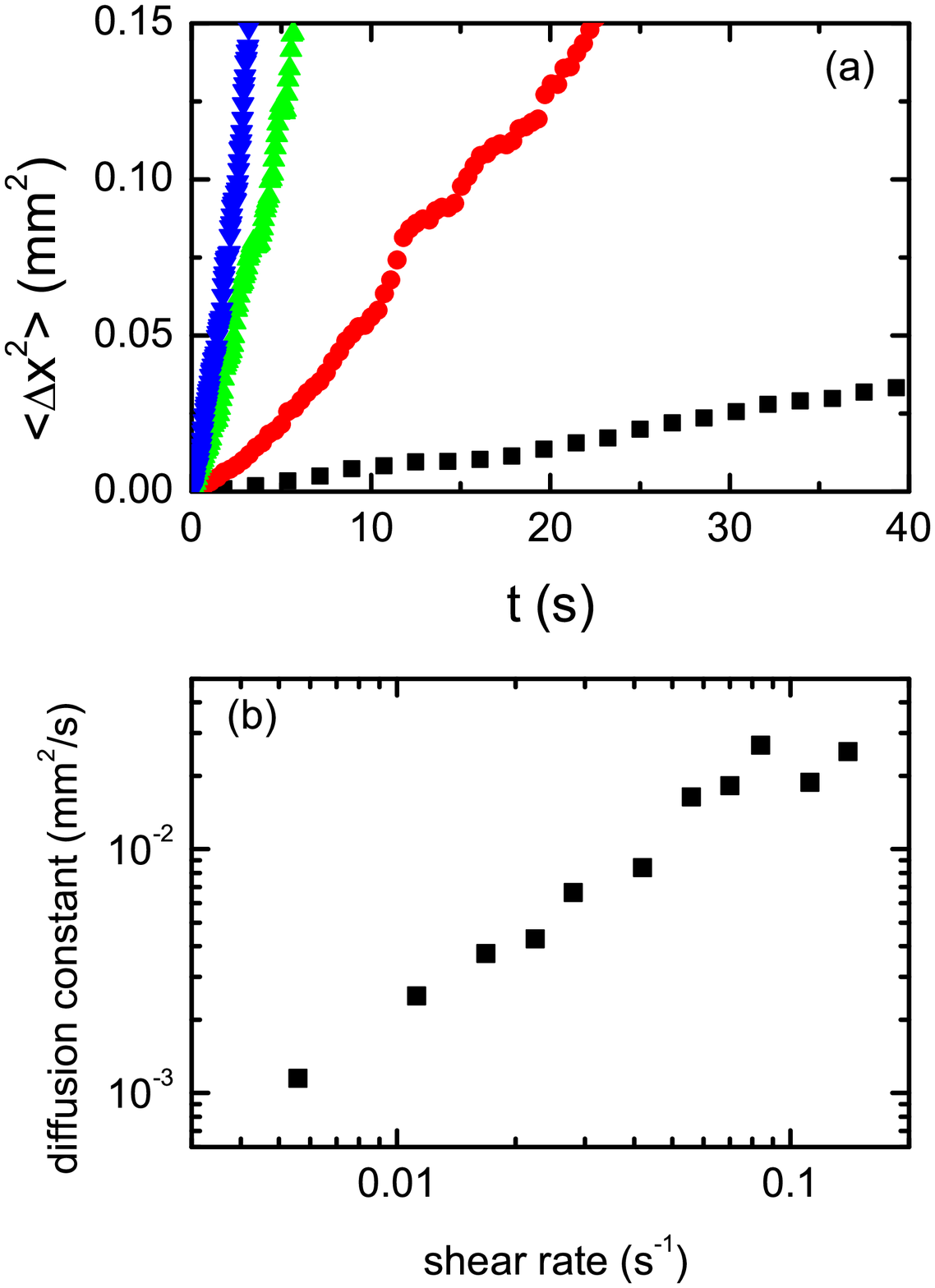}
\caption{ (color online) (a) The $<(\Delta x(t))^2>$ averaged over
all the bubbles as a function of time for bubbles from the central
bin at four different rates of strain: $0.0028\ {\rm s^{-1}}$
($\blacksquare$); $0.014\ {\rm s^{-1}}$ (red $\bullet$); $0.07\
{\rm s^{-1}}$ (green $\blacktriangle$); and $0.14\ {\rm s^{-1}}$
(blue $\blacktriangledown$). (b)  Diffusion constants for motion
transverse to the direction of flow as a function of rate of
strain. The diffusion constants are based on the behavior of
bubbles in the central bin, and they are seen to scale as a
power-law with the rate of strain.} \label{diffusion constants}
\end{figure}

As a first approximation, we characterize $<(\Delta x(t))^2>$ as
growing linearly with time. We can use this to estimate the
effective diffusion constant for individual bubbles along the
y-direction. Figure~\ref{diffusion constants}a is a plot of
$<(\Delta x(t))^2>$ for a range of rates of strain for bubbles in
the central bin. The slope of these curves yield the diffusion
constant, and this is plotted in Fig.~\ref{diffusion constants}b.
Unfortunately, the range of rate of strain that we were able to
access is relatively small in comparison to numerical simulations
as in \cite{OTLL03}. In this limited range, the diffusion constant
shows a linear dependence on the rate of strain.

\begin{figure}
\includegraphics[width=8cm]{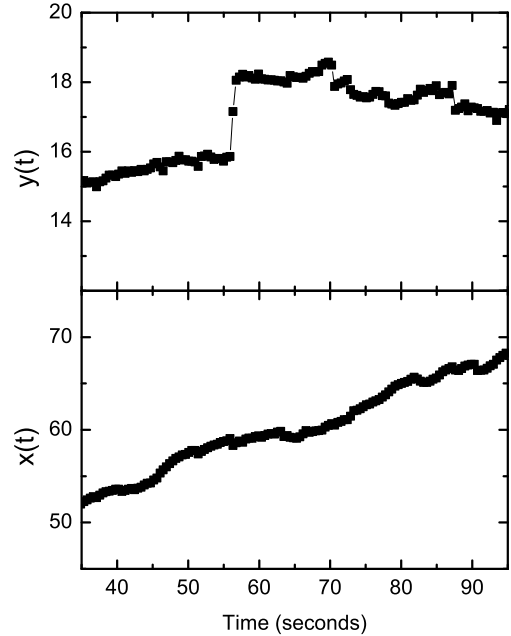}
\caption{The above plots indicate the x- and y-displacements of a
single bubble with time. The fluctuations are qualitatively
different, with ballistic-like transport dominating the shear
direction and motion similar to Levi-flights in the transverse
direction.} \label{teaser}
\end{figure}

\section{Discussion}

We have performed an extensive investigation of kinematics
associated with a sheared bubble raft. A striking feature of the
results is the nature of the probability distributions of the
velocity. For the distributions in the central region of the
trough, they are well-fit by Lorentzian functions. This is very
different from what one would expect for a thermal distribution of
velocities. At this point, more work is needed on the individual
bubble kinematics to determine the source of this distribution.
But, a potential candidate is the highly overdamped dynamics.

The other feature of the distributions that requires further study
is the asymmetry that develops in the probability distribution for
$v_x$ when bubbles that are off-center are considered. One
potential explanation is the influence of the boundaries. However,
there does not appear to be any similar asymmetry in the
distribution for $v_y$. This suggests that the boundaries are not
the source of the asymmetry, as one might expect the boundaries to
influence both $v_x$ and $v_y$. A related issue is the slightly
nonlinear behavior of the average value of $v_x$. The source of
the nonlinearity and the asymmetry of the probability distribution
may be connected. Further work will be done in this area.

When summarizing the results of these experiments, it is
especially helpful to compare and contrast them with numerical
studies of the bubble model that have guided earlier
investigations. Not surprisingly, the general qualitative behavior
matches the bubble model. A key element of the bubble model is the
existence of a critical rate of strain \cite{OTLL03} that is
identified by a changes in the behavior of various quantities. One
quantity in particular is the probability distribution of $v_y$.
In the bubble model, this distribution develops a significant flat
region for small values of $v_y$ above a critical rate. We did not
observe this behavior, and the measured distributions of $v_y$ all
suggest that the rate of strains studied in this paper are below
the critical rate of strain.

Another measurement for which we observed reasonable qualitative
agreement between the experiment and the bubble model simulations
is in regard to the diffusion constant. However, we are unable to
make a number of quantitative comparisons. First, it would be
useful to study a wider range of rates of strain, particularly
higher ones. However, currently we are limited in our studies of
diffusion at higher rates of strain due to bubble lifetimes and
the fact that they are eventually swept out of the system. Similar
issues prevented calculating the diffusion constant from velocity
correlation functions. These measurements will require either a
longer system or the use of a Couette geometry which mimics
periodic boundary conditions.

Despite the generally good qualitative agreement, there are some
disagreements between our measurements and the bubble model. For
example, the exponent for the power-law scaling of the width of
distribution for velocity fluctuations is different for the bubble
model studies reported in Ref.~\cite{OTLL03} and this current
work. However, it should be noted that other work on fluctuations
in a Couette geometry \cite{D05} were consistent with the work of
Ref.~\cite{OTLL03}. When comparing the current work to
Ref.~\cite{D05}, there are two main differences: geometry and
degree of polydispersity. As the geometry is the same between this
work and the simulations of the bubble model, it is most likely
that the exact value of the exponent for the scaling of
fluctuations is sensitive to the polydispersity. Future work will
be able to test this systematically.

A more significant departure from results in the bubble model
\cite{OTLL03} and our measurements is the existence of a peak in
the $\Delta v_x$ distribution. More experimental work is needed to
determine the source of this peak. Two obvious candidates are the
the asymmetry of the velocity distributions or a characteristic
associated with the details of the bubble rearrangements. As the
asymmetry in the velocity distribution near the boundaries is
unique to our experimental study, this is definitely an effect
that is not captured by the bubble model and could explain the
lack of a peak in the velocity distribution. However, it is also
possible that the details of the bubble rearrangements differ
between the bubble rafts and the bubble model. This is one
motivation for a closer look at individual bubble kinematics in
the experiments.

One element missing from the bubble model is the attractions
between the bubbles. This is an obvious candidate as the source of
the quantitative disagreements between the bubble model
simulations and our experiments. Incorporating such attractive
effects in numerical models may help remove discrepancies with
experiments.

For a number of the open questions, we have commented on the need
for more detailed studies of the individual bubble dynamics. As
useful as the collective studies reported in this paper, studies
of individual particle kinematics provide insights into mechanisms
of the microscopic transport taking place. Figure~\ref{teaser}
illustrates the type of behavior that requires further study. In
Fig.~\ref{teaser}, we note the different qualitative behavior in
transport transverse and longitudinal to the shear for a single
bubble. There is an almost linear dependence of displacement with
time along the x-direction due to the imposed flow. In the
y-direction, the bubbles experience periods of extremely small
displacements punctuated with large fluctuations. These
fluctuations are of the length scale of about a bubble diameter,
and may be attributed to the occurrence of T1 events in the
immediate neighborhood of the bubble. It will be interesting to
connect this behavior to similar motions in glassy systems
\cite{WW02,DDKPPG98}, and we are currently developing measures
characterizing this qualitatively different transport.

\begin{acknowledgments}

This work was supported by a Department of Energy grant
DE-FG02-03ED46071.

\end{acknowledgments}


\end{document}